\documentclass{article}
\usepackage{spconf,amsmath,graphicx}

\usepackage[misc, geometry]{ifsym}
\usepackage{hyperref}
\hypersetup{
    colorlinks=true,
    linkcolor=blue,
    filecolor=magenta,      
    urlcolor=cyan,
}
\usepackage{enumitem}
\setlist{nosep, leftmargin=14pt}

\usepackage{mwe} 

\newcommand{\ve}[1]{\mathbf{#1}}

\title{Towards Unbiased COVID-19 Lesion Localisation and Segmentation via Weakly Supervised Learning}
\name{
\begin{tabular}{@{}c@{}}
\qquad Yang Yang$^1$$^,$$^2$ \qquad Jiancong Chen$^1$  \qquad Ruixuan Wang$^2$$^*$   \qquad Ting Ma$^1$$^,$$^3$   \qquad Lingwei Wang$^4$ \\
Jie Chen$^1$$^,$$^5$ \qquad Wei-Shi Zheng$^2$  \qquad Tong Zhang$^1$$^*$\thanks{$^*$Coresponding authors}
\thanks{Email-to:  zhangt02@pcl.ac.cn, wangruix5@mail.sysu.edu.cn } 
\end{tabular}}
\address{$^1$ Artificial Intelligence Research Center, Peng Cheng Laboratory, Shenzhen, China\\
$^2$ School of Computer Science and Engineering, Sun Yat-sen University, China~\\
$^3$School of Electronic and Information Engineering, Harbin Institute of Technology, Shenzhen, China~\\
$^4$Shenzhen People's Hospital, Shenzhen Institute of Respiratory Diseases, Shenzhen, China~\\
$^5$School of Electronic and Computer Engineering, Peking University Shenzhen Graduate School, China}

\begin{document}

\maketitle

\begin{abstract}
Despite tremendous efforts, it is very challenging to generate a robust model to assist in the accurate quantification assessment of COVID-19 on chest CT images. Due to the nature of blurred boundaries, the supervised segmentation methods usually suffer from annotation biases. To support unbiased lesion localisation and to minimise the labelling costs, we propose a data-driven  framework supervised by only image level labels. The framework can explicitly separate potential lesions from original images, with the help of an generative adversarial network and a lesion-specific decoder. Experiments on two COVID-19 datasets demonstrates the effectiveness of the proposed framework and its superior performance to several existing methods.
\end{abstract}

%
%
\section{Introduction}
\label{sec:intro}
Since the first case reported in Dec 2019, the novel Coronavirus Disease (COVID-19) has made the world a pandemic era. Till 4 Oct 2020, there have been 34,724,785 confirmed cases of COVID-19, including 1,030,160 deaths, according to WHO~\cite{wang2020novel}. Accurate lesion localisation and segmentation methods are in huge demand to aid the fast disease diagnosis and stage monitoring. Among different diagnostic imaging modalities, computed tomography (CT) has proven itself to be effective and been widely used for the assessment and evaluation of disease evolution~\cite{zhang2020clinically,rubin2020role}. Patchy ground-glass opacitity (GGO) with consolidation is often been found from CT images as a typical sign of lung infection. Thus, the quantitative evaluation of such lung lesions can help  diagnosis. 

Recently, deep learning algorithms, e.g., Convolutional Neural Networks (CNNs)~\cite{yasaka2018deep}, have been widely used to detect lung diseases via CT images. For example, researchers applied existing CNN frameworks, such as U-Net~\cite{ronneberger2015u}, to the automatic segmentation of COVID-19 CT scans~\cite{huang2020serial,li2020artificial,cao2020longitudinal}. To achieve satisfactory results, the highly accurate annotation of lesions is essential. However, obtaining a large amount of annotation of infections is expensive and time-consuming. 
COPLE-Net was designed to enhance the robustness of the detection, using the labels polluted by the noise data from non-experts~\cite{wang2020noise}. 
Another ways is to use a weakly supervised framework for classification and localization of lesions~\cite{wang2020weakly}. Yet, it is still difficult for these methods to identify the boundaries of GGO as a result of its low contrast and blurred appearances.

To overcome above challenges, we propose a novel weakly supervised framework for automatic localization and segmentation of COVID-19 pneumonia lesions only with the help of image-level label information. The framework consists of a generative adversarial network and an additional decoder specifically for lesion estimation. It can  explicitly decompose any image into two images, one containing the normal information in the original image, and the other containing possible lesion information if existing in the original image. An effective training strategy with new loss terms was proposed to help decompose potential lesions from normal information in images. Extensive evaluations (including cross-dataset evaluation) on two COVID-19 datasets confirmed the effectiveness of the proposed method in lesion localization and segmentation.

\begin{figure*}[!btp]
    \centering
    \includegraphics[width=0.7\textwidth]{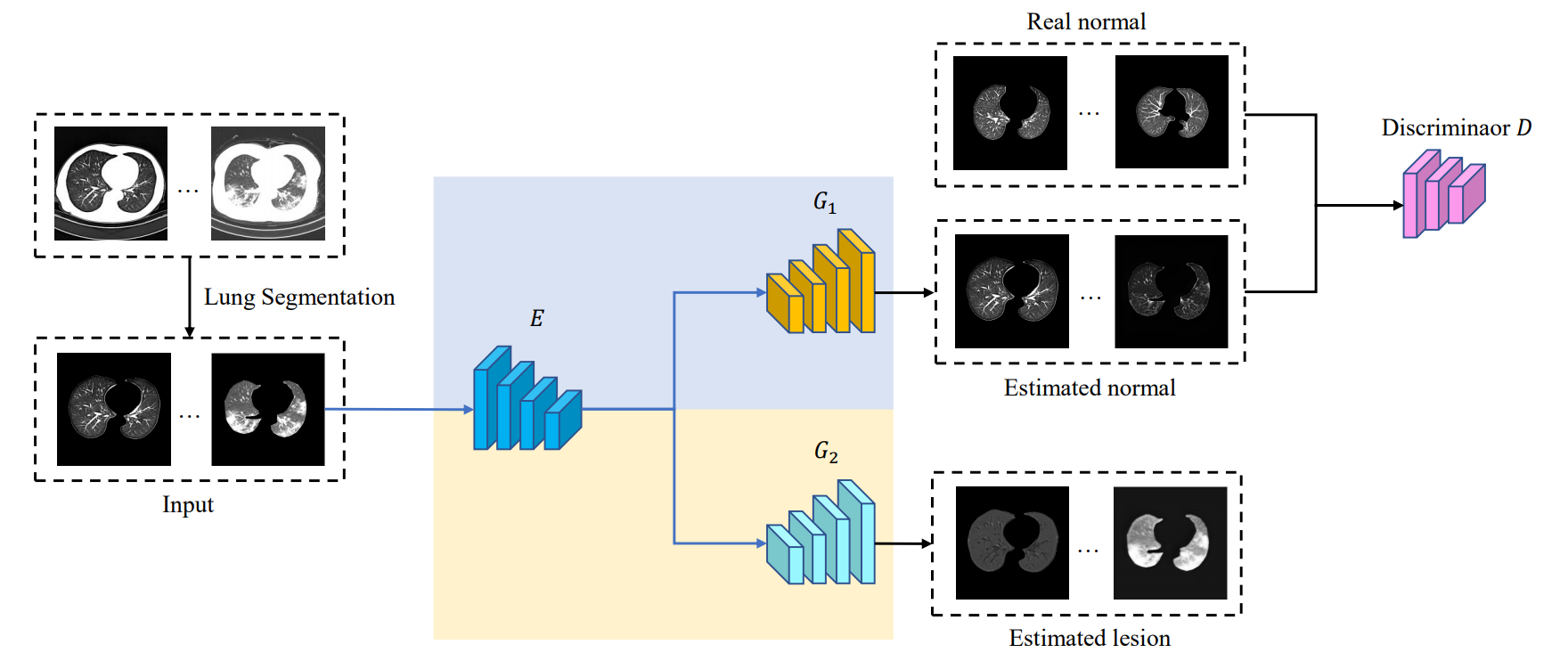}
    \caption{The proposed weakly supervised framework for lesion localization and segmentation. It consists of the encoder $E$, the generator $G_1$ for estimation of normal information from the input, the decoder $G_2$ for estimation of lesion information from the input, and the discriminator (critic) $D$ to judge whether the generator's outputs are realistically normal or not. A lung segmentation model was pre-trained and applied to the original CT slices before they are input to the network model.}
    \label{fig:architecture}
\end{figure*}

\section{Method}
\label{sec:mothod}

\subsection{Overview of the framework}
We propose a novel weakly supervised framework for automatic localization and segmentation of COVID-19 pneumonia lesions only with the help of image-level label information (Figure~\ref{fig:architecture}). The basic idea is to explicitly decompose any image (either normal or with lesion) into a corresponding normal version and a remaining lesion version, with the constraint that there should be no lesion in the lesion version for any normal image. To help obtain realistic normal versions from lesioned images, a discriminator $D$ is employed to judge whether the decomposed normal versions are realistic or not compared to real normal images. Therefore the framework can be considered as the fusion of a generative adversarial network (GAN) and a lesion decoder, with the lesion decoder part $G_2$ sharing the same encoder $E$ with the generator $G_1$ (Figure~\ref{fig:architecture}).

\begin{figure*}[!bth]
    \centering
    \includegraphics[width=1\linewidth]{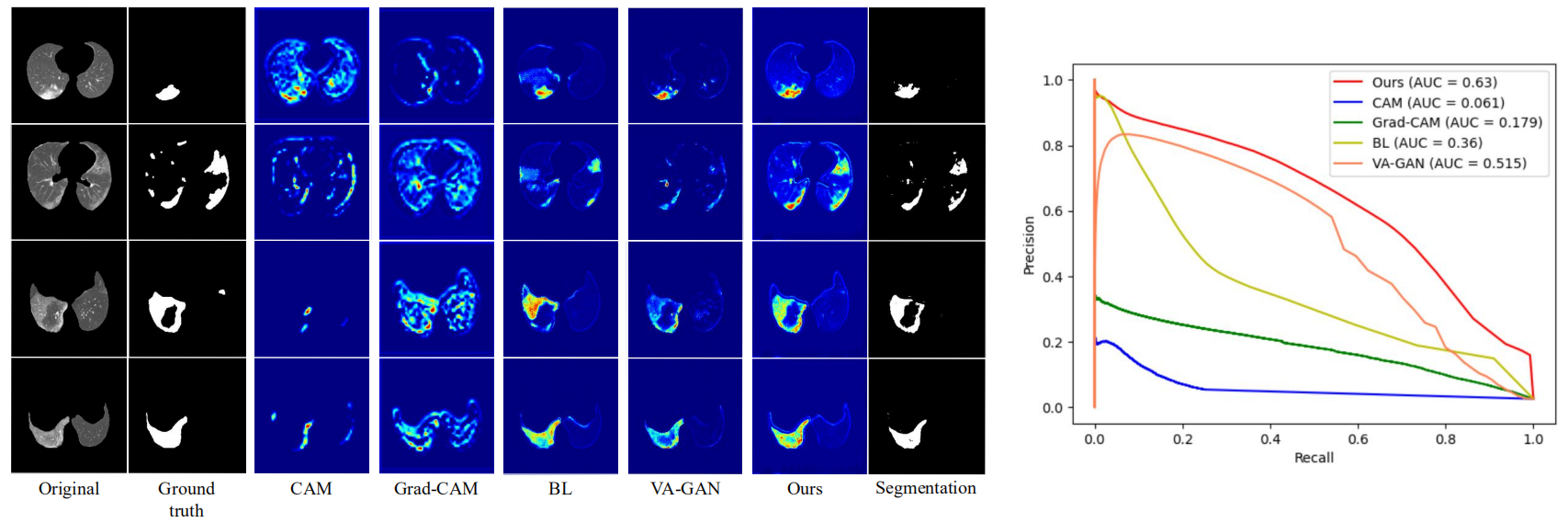}
    \caption{Comparison with baseline methods on the  COVID-19  Image  Data  Collection dataset. Left (in column): input images, ground-truth lesion area, localization results by CAM, Grad-CAM, BL, VA-GAN, our approach and segmentation of our approach (with threshold 0.4). Red regions indicate higher probabilities to be lesion. Right: PR curve for each method.}
    \label{fig:compare}
\end{figure*}

\subsection{Model training}

Suppose a set of lung CT slices $\mathcal{D}=\{(\ve{x}_i, y_i)\}_{i=1}^N$ with lung regions of interest (ROI) pre-segmented are available to train the network model, where $\ve{x}_i$ denotes the $i$-th CT slice and $y_i$ denotes whether the slice is normal ($y_i=0$) or contains lesion ($y_i=1$). For any slice image $\ve{x}_i$ as the input to the model, denote by $G_1(E(\ve{x}_i))$ the output of the generator $G_1$, representing the normal version of the original input $\ve{x}_i$, and by $G_2(E(\ve{x}_i))$ the output of the decoder $G_2$, representing the lesion information in the original input $\ve{x}_i$. If the decomposition process works well, the recombination of the two decomposed components should be close to the original input, i.e., the reconstruction loss $L_r$ should be small,
\begin{equation} \label{eq:reconstruct_loss}
L_r = \frac{1}{N}\sum_{i=1}^N\left\|\ve{x}_i - G_1(E(\ve{x}_i)) -G_2(E(\ve{x}_i)) \right\|  \,  
\end{equation}
where $\|\cdot\|$ represents the $L_p$ norm with $p=1$ or $2$.
Since normal images contain no lesion, if the decomposition works well, the normal version $G_1(E(\ve{x}_j))$ itself should be close to the original input for any normal input $\ve{x}_j$, i.e., the normal fidelity loss $L_g$ for normal images should be small, 
\begin{equation} \label{eq:reconstruct_loss}
L_n = \frac{1}{N_1}\sum_{j=1}^{N_1}\left\|\ve{x}_j - G_1(E(\ve{x}_j)) \right\|  \,  
\end{equation}
where $\ve{x}_j$ is the $j$-th normal image and $N_1$ is the total number of normal images. The set of normal images is a subset of the whole dataset $\mathcal{D}$. 

While the minimization of $L_r$ and $L_g$ together may help the model well reconstruct normal images, it may not be enough to correctly estimate the lesion information by the decoder  $G_2(E(\ve{x}_i))$ when the input image $\ve{x}_i$ contains lesion, because there could exist multiple or even infinite number of decomposition results which can satisfy the constraint $\ve{x}_i = G_1(E(\ve{x}_i)) + G_2(E(\ve{x}_i))$, i.e., making $L_r$ minimal. An extreme case is that the generator $G_1$ would always output the original input, no matter whether the input contains lesion or not, which would make the lesion decoder output little or no information about lesion. To well separate lesion from healthy parts in lesioned images, the proposed framework uses a discriminator to judge whether the decomposed normal versions $G_1(E(\ve{x}_i))$'s  are really similar to real normal images or not. Here the Wasserstain GAN with gradient penalty (WGAN-GP) is adopted to train the discriminator (also called critic) $D$. Recall that, in the current task, the sampling version of the loss function for the critic of WGAN-GP is 
\begin{equation}\label{eq:wgangp}
L_c = \frac{1}{N}\sum_{i=1}^N D(G_1(E(\ve{x}_i))) -
\frac{1}{N_1}\sum_{j=1}^{N_1} D(\ve{x}_j) + 
 \lambda \cdot GP \,
\end{equation}
where $GP$ stands for the gradient penalty term (see detailed form in \cite{gulrajani2017improved}) and $\lambda$ is its corresponding weight. This loss aims at maximizing the critic output for real normal data $\ve{x}_j$'s, meanwhile minimizing the critic output for estimated normal data $G_1(E(\ve{x}_i))$'s from the generator $G_1$. Higher output indicates that the input to the critic is more realistic. On the other hand, as part of the well-known alternative GAN training strategy, the generator $G_1$ together with the encoder $E$ can be trained by minimizing the adversarial loss $L_a$,
 \begin{equation}
 L_a = - \frac{1}{N}\sum_{i=1}^N D(G_1(E(\ve{x}_i))) \,
 \end{equation}
Minimization of this loss would help the generator output more realistic normal estimates, resulting in higher output from the critic $D$.

Overall, the generator $G_1$, the encoder $E$, and the lesion decoder $G_2$ can be trained together by minimizing the combined loss terms $L_r$, $L_g$ and $L_c$, 
 \begin{equation}
\label{eq:overallg}
L_g = \alpha_1 L_a + \alpha_2 L_r +\alpha_3 L_n \,
\end{equation}
$\alpha_1$, $\alpha_2$, and $\alpha_3$ are coefficients to trade off the importance between the three loss terms. $L_c$ and $L_g$ are minimized alternatively to train the critic (discriminator) and the other parts of the network model.
\section{Experimental Evaluation}
\label{sec:experiment}

\subsection{Experimental settings}
The proposed network was trained with a set of 2007 normal lung CT slices and 870 lesioned slices which were randomly sampled from the COVID-cell dataset~\cite{zhang2020clinically}, and then evaluated on two test sets of lesioned images. One test set includes 128 lesioned images where were randomly sampled from the COVID-cell dataset and then annotated at pixel level by one logists. Note that there is no overlap between the training set and this test set although they are both from the COVID-cell dataset.
The other test set consist of 493 lesioned slices from the COVID-19 Image Data Collection~\cite{cohen2020covid} which contains 20 cases with covid-19. This dataset was released with lesion area annotations, although the pixel-level annotations are not that accurate particularly around the  boundary of the lesion area. 
It is worth noting that all the pixel-level lesion annotations were not for model training but only for quantitative evaluation of the proposed model.

As a pre-processing step, the lung regions in all images in both training and test sets were segmented out with a U-Net segmentation model, where the segmentation model was pre-trained on covid-cell dataset. The visual information outside the lung region was removed from each image based on the segmentation mask before the image was used for training or testing. Each image was resized to $256\times 256$ pixels, and then normalized based on the mean and standard deviation of pixel values over all training images.
\begin{figure}[!bth]
    \centering
    \includegraphics[width=0.9\linewidth]{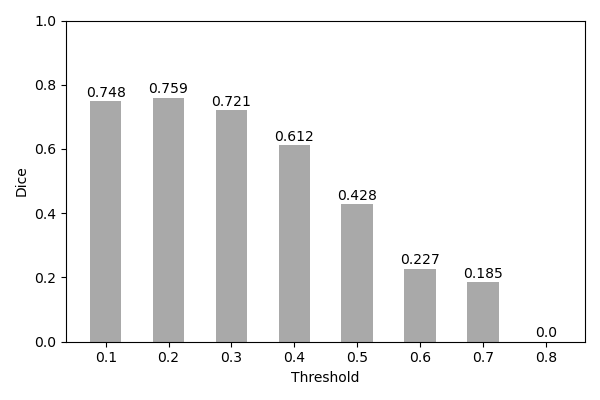}
    \caption{Segmentation performance (dice score) on the COVID-cell dataset, with different thresholds on heatmaps.}
    \label{fig:dice_score}
\end{figure}

\begin{figure*}[!bthp]
    \centering
    \includegraphics[width=1\linewidth]{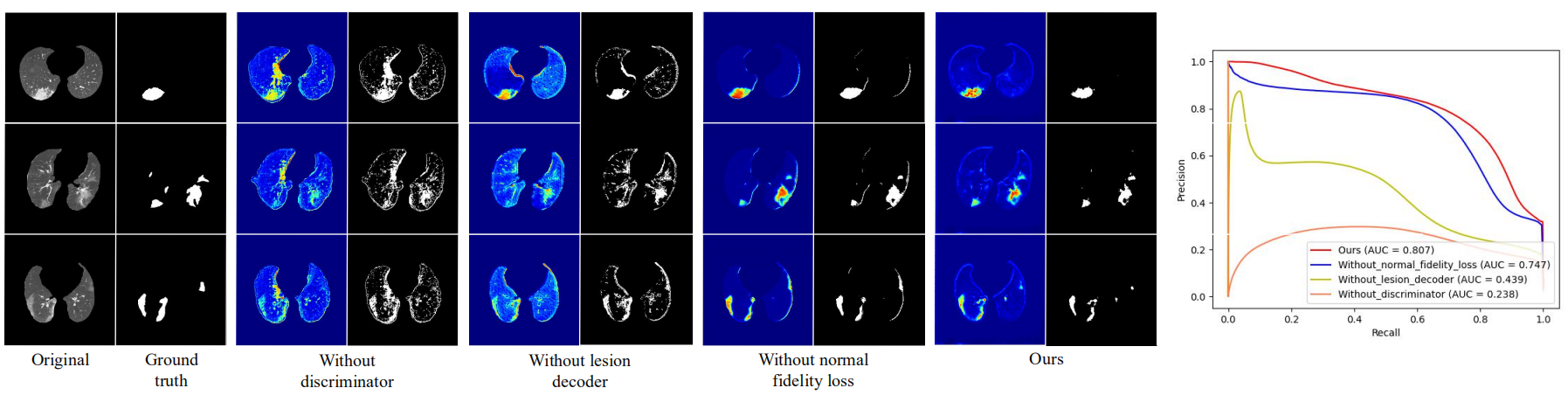}
    \caption{Ablation study on COVID-cell dataset. Left (in column): input images, ground-truth lesions, our approach without normal fidelity loss, without lesion decoder, without discriminator, and our approach. Segmentation is based on threshold 0.4 on the heatmaps. Right: PR curve for each condition.}
    \label{fig:ablation}
\end{figure*}

In our proposed framework, the encoder $E$ and the generator $G_1$ forms the well-known U-Net network, and similarly the 
encoder $E$ and the lesion decoder $G_2$ forms the other U-Net network. The only modification is the addition of the Tanh activation function at the last layer of the generator $G_1$ and the decoder $G_2$ respectively to constrain the pixel values of the output within the same range $(-1, 1)$ as that of the model's input. A seven-layer CNN was used for the discriminator (or critic) $D$, with the outputs of three down-samplings pooled globally and then concatenated to form the input to the final fully connected layer. For model training, the gradient penalty coefficient $\lambda$ in the WGAN loss was empirically set to 10, and  the coefficients $\alpha_1=0.01$, $\alpha_2=\alpha_3=100$.
Adam was adopted as the optimizer during model training, with default learning rate 0.0002, and batch size 8.  The PR curves are used to evaluate localization performance of the proposed model and baseline methods, which were generated by comparing the pixel-level lesion estimates (from the output of the decoder $G_2$) with ground truth annotations. The output values from the decoder was normalized from [-1, 1] to [0, 1] for both quantitative and qualitative evaluations.  


\subsection{Comparison with other methods}

We compared the lesion localization ability of our method with the commonly used visualization techniques CAM~\cite{zhou2016learning}, Grad-CAM~\cite{selvaraju2017grad}, and the recent visual feature attribution using Wasserstein GANs $($VA-GAN$)$~\cite{baumgartner2018visual}, and biomarker localization (BL) method~\cite{zhang2019biomarker}. ResNet18 was chosen as the backbone of CAM and Grad-CAM to train a binary classification task. From Figure~\ref{fig:compare} (Left), it can be observed that CAM fails to detect most lesion areas and Grad-CAM detects the lesions but introduces many irrelevant area. 
As for BL, it can indeed localize some lesions, but often fails to detect small or indistinct lesions. In comparison, our method provides much more precise localization (and therefore segmentation) of lesions even if the lesions are in irregular shapes or with vague boundaries, demonstrating the superior performance of our method to others. 
This is also confirmed by the quantitative evaluation in the PR curve (Figure~\ref{fig:compare}, Right), with the area under the PR curve (AUC) 0.63 for our approach, 0.36 for BL, 0.179 for Grad-CAM and 0.061 for CAM.
Based on the localization results, lesions can be automatically segmented by thresholding the heatmaps. Figure~\ref{fig:dice_score} shows that in a wide range of thresholds ([0.1, 0.4]), the simple threshold-based segmentation resulted in the dice score above 0.7, suggesting that the proposed framework can provide reasonably good segmentation even just based on the image-level labels.
\subsection{Ablation study}

In this section, we evaluate the effect of loss terms and different components in our framework by ablation study. 
As can be seen from Figure~\ref{fig:ablation} (Left), with the removal of different framework components or part of loss terms, the localization and corresponding segmentation (threshold=0.4) performance degrades more or less. In particular, by removing the lesion decoder or the discriminator, the model fails to discriminate the lesion regions from many normal regions. Without the normal fidelity loss, some normal boundaries were mistakenly considered as lesions by the model. Quantitative evaluation (Figure~\ref{fig:ablation}, Right) further confirmed the effectiveness of each component or loss term in performance boosting.

\section{Conclusion}
\label{sec:conclusion}

In this paper, an effective weakly supervised localization and segmentation framework is proposed.  Experiments on two lung CT datasets demonstrate that the proposed framework achieves superior performance compared with widely used visualization methods and a recent lesion localization method. Without annotating the detailed lesion regions, the proposed framework
provides a novel and effect approach for clinicians to efficiently analyze degree of lesions based on the automatic localization and segmentation results, particularly for the diagnosis of COVID-19 disease. Source codes implemented in PyTorch and MindSpore will be available at \url{https://git.pcl.ac.cn/capepoint} after the conference.

\subsection*{Compliance with Ethical Standards}
This work proposes a general machine learning framework for weakly supervised image localisation and segmentation. Our experimental data were collected from open source datasets, which were ethically approved as indicated in the references~\cite{zhang2020clinically, cohen2020covid}.
\subsection*{Acknowledgments}
This work is supported by the Guangdong Provincial Science and Technology Department (grant No. 2020B1111340056), the National Natural Science Foundation of China (grant No. 62071502, U1811461, 61972217, 62081360152), the Guangdong Key Research and Development Program (grant No. 2019B020228001, 2020B1111190001), Guangdong Basic and Applied Basic Research Foundation (grant No. 2019B1515120049), and the Meizhou Science and Technology Program (grant No. 2019A0102005). We also thank Major Empowerment for Artificial Intelligence Research Project from Peng Cheng Laboratory.

\bibliographystyle{IEEEbib}
\bibliography{refs}

\end{document}